\documentclass[conference]{IEEEtran}
\usepackage[utf8]{inputenc}

\usepackage{xspace}
\usepackage{url}
\usepackage{graphicx}
\usepackage{epstopdf}
\usepackage{paralist}
\usepackage[normalem]{ulem}
\usepackage{fancyref} 

\usepackage{amsmath}
\usepackage{amsfonts}
\usepackage{amssymb}

\usepackage{bm}

\usepackage{textcomp}
\usepackage{array}
\usepackage{arydshln}
\usepackage{mwe}
\usepackage{url}
\usepackage{siunitx} 
\usepackage{tikz}
\usepackage{gensymb}

\usepackage[inline]{enumitem}
\usepackage{xcolor}
\definecolor{links}{rgb}{0.3,0,0}   
\definecolor{urls}{rgb}{0,0,0.8}    
\definecolor{cites}{rgb}{0,0,0.6}   

\definecolor{silver}{cmyk}{0,0,0,0.3}
\definecolor{navy}{cmyk}{0.8,0.5,0,0}
\definecolor{lightblue}{cmyk}{0.35,0.11,0,0}
\definecolor{orange}{cmyk}{0,0.57,0.86,0}
\definecolor{yellow}{cmyk}{0,0,0.9,0.0}
\definecolor{reddishyellow}{cmyk}{0,0.22,1.0,0.0}
\definecolor{lightred}{cmyk}{0,0.820,0.753,0.0}
\definecolor{black}{cmyk}{0,0,0.0,1.0}
\definecolor{white}{cmyk}{0,0,0.0,0}
\definecolor{purple}{cmyk}{0.64,0.83,0,0}
\definecolor{darkYellow}{cmyk}{0,0,1.0,0.5}
\definecolor{darkSilver}{cmyk}{0,0,0,0.1}
\definecolor{lightyellow}{cmyk}{0,0,0.3,0.0}
\definecolor{lighteryellow}{cmyk}{0,0,0.1,0.0}
\definecolor{lightestyellow}{cmyk}{0,0,0.05,0.0}
\definecolor{darkblue}{cmyk}{0.98,0.89,0,0.11}
\definecolor{bluel1}{cmyk}{0.5,0.05,0.05,0.05}
\definecolor{darkred}{cmyk}{0,0.89,0.7,0.55}
\definecolor{magenta}{rgb}{1.0, 0.0, 1.0}
\definecolor{cyan}{rgb}{0.0, 1.0, 1.0}
\usepackage{dsfont}
\usepackage{footnote}
\usepackage{balance}
\usepackage{float}
\usepackage{cite}

\usepackage{wasysym}
\usepackage{amssymb}
\usepackage{tabularx,booktabs}
\newcolumntype{L}[1]{>{\raggedright\let\newline\\\arraybackslash\hspace{0pt}}m{#1}}
\newcolumntype{C}[1]{>{\centering\let\newline\\\arraybackslash\hspace{0pt}}m{#1}}
\newcolumntype{R}[1]{>{\raggedleft\let\newline\\\arraybackslash\hspace{0pt}}m{#1}}
\usepackage{boldline}

\usepackage{algorithmicx}
\usepackage{algorithm}
\usepackage{algpseudocode}
\algnewcommand\ForGiven{\textbf{for a given }}
\algnewcommand\algorithmiccompute{\textbf{Compute }}
\algnewcommand\Compute{\item[\algorithmiccompute]}

\usepackage{subcaption}
\usepackage{pgfplots}
\usepgfplotslibrary{groupplots}
\usetikzlibrary{pgfplots.groupplots}

\usepackage{vmr-symbols-vecbold}
\usepackage{standard-macros}

\newcommand{\rate}{\mathrm{R}}
\newcommand{\mN}{\mathrm{N}}

\newcommand{\mM}{\mathrm{M}}

\newcommand{\mT}{\mathrm{T}}

\newcommand{\mP}{\mathrm{P}}

\let\nd\undefined
\let\ng\undefined
\newcommand{\nb}{n_{\text{b}}}
\newcommand{\nc}{n_{\text{c}} }
\newcommand{\np}{n_{\text{p}} }
\newcommand{\nd}{n_{\text{d}} }
\newcommand{\ng}{n_{\text{g}} }

\newcommand{\infden}{\imath_{s}}

\safemath{\tR}{\tilde{\rate}}
\safemath{\tgamma}{\tilde{\gamma}}
\safemath{\mfn}{\mathfrak{N}_{n,\gamma''(\zeta)}}
\safemath{\mfnt}{\tilde{\mathfrak{N}}_{n,\gamma''(\zeta)}}

\begin{document}
\voffset = 0.02in
\title{Pilot-Assisted Faster-than-Nyquist Signaling for HRLLC: A Non-Asymptotic Approach}

\author{\IEEEauthorblockN{A. Oguz Kislal\textsuperscript{1}, Melda Yuksel\textsuperscript{2}, Ali Gorcin\textsuperscript{1,3},  Halim Yanikomeroglu\textsuperscript{4}}
\IEEEauthorblockA{\textsuperscript{1}\textit{Communications and Signal Processing Research (HISAR) Lab, TUBITAK BILGEM, Kocaeli, Turkiye}\\
\textsuperscript{2}\textit{Department of Electrical and Electronics Engineering, Middle East Technical University, Ankara, Turkey} \\
\textsuperscript{3}\textit{Faculty of Electrical and Electronics Engineering, Istanbul Technical University, Istanbul, Turkiye} \\ 
\textsuperscript{4}\textit{Non-Terrestrial Networks (NTN) Lab, Systems and Computer Engineering, Carleton University, Ottawa, ON, Canada} \\
Emails: ahmet.kislal@tubitak.gov.tr, ymelda@metu.edu.tr, aligorcin@itu.edu.tr, halim@sce.carleton.ca}
}
\maketitle
\begin{abstract}
This paper investigates the performance of faster-than-Nyquist (FTN) signaling within the context of hyper-reliable low-latency communications (HRLLC), specifically focusing on the challenges imposed by the short-packet regime. While traditional Nyquist-based systems maintain symbol orthogonality to prevent inter-symbol interference (ISI), FTN intentionally introduces ISI to achieve higher transmission rates. While many existing FTN studies assume perfect channel state information, this assumption is often impractical for mission-critical HRLLC. In such scenarios, a portion of the limited packet length must be reserved for pilot symbols to ensure reliable estimation. To characterize the achievable error probability while accounting for imperfect channel estimation in the short-blocklength regime, we derive the random coding union bound with parameter $s$ (RCUs) under mismatched decoding for FTN systems. The numerical results demonstrate that FTN provides up to a $2 \dB$ SNR gain over Nyquist signaling, provided that power allocation and pilot overhead are optimized. These findings highlight the necessity of non-asymptotic analysis for designing efficient, next-generation HRLLC-FTN systems.
\end{abstract}

\begin{IEEEkeywords}
Faster-than-Nyquist, finite-blocklength information theory, pilot-assisted transmission, hyper-reliable low-latency communication.
\end{IEEEkeywords}

\section{Introduction}
Modern wireless communication systems, specifically the emerging 6G architecture and its integration with next-generation networks, are designed to support mission-critical links that must satisfy stringent reliability and latency requirements. Known as hyper-reliable low-latency communications (HRLLC)~\cite{Yiyang2026}, this capability is a cornerstone for enabling transformative applications such as vehicle-to-everything (V2X) communications~\cite{Ince2024}, factory automation~\cite{Liu2025}, autonomous driving~\cite{Lin2026} and haptic communications~\cite{Tao2023}. 

A defining trait of HRLLC is the prevalence of compact information payloads delivered through short packets, which is a direct consequence of the physical constraints on data transmission as the product of signal duration and bandwidth determines packet length. In HRLLC scenarios, latency requirements strictly limit the signal duration, while bandwidth is frequently partitioned to prevent multiuser interference that would otherwise degrade reliability. Consequently, standard asymptotic metrics such as ergodic capacity and outage rates fail to accurately model performance in this regime~\cite{durisi16-09a}. In addition, these metrics do not account for the potential gains from channel diversity, nor for the unavoidable overhead associated with pilot-based channel estimation~\cite{Johan2019}. Although exploiting diversity can boost reliability in short blocks, the symbols sacrificed for training signals significantly reduce the effective data rate. Therefore, a rigorous, non-asymptotic characterization of the trade-off between transmission rate and packet error probability is needed.

Faster-than-Nyquist (FTN) transmission is an emerging technology that increases data rates by intentionally violating the Nyquist criterion \cite{Mazo1975}. While traditional systems strictly enforce orthogonality of symbols to prevent inter-symbol interference (ISI), FTN intentionally introduces controlled ISI \cite{Anderson2013}. By managing this controlled interference, FTN  achieves transmission rates that are not unachievable under standard Nyquist constraints \cite{Ganji2020,Zhang2023,Zichao2023}.

Although extensive research exists on achievable coding rates for Nyquist transmission under finite-blocklength constraints \cite{Polyanskiy2010}, only a limited number of studies analyze the performance of FTN in the finite-blocklength regime \cite{Mohammadkarimi2021,Zhang2025Mar,Kim2026}. In \cite{Mohammadkarimi2021}, the maximum rate for a given blocklength and packet error probability for a SISO FTN system is derived under an input power constraint. In \cite{Zhang2025Mar} these results are extended to a MIMO FTN system. 
In \cite{Kim2026}, the analyses in \cite{Mohammadkarimi2021} and \cite{Zhang2025Mar} are further extended to a general framework utilizing the time-bandwidth product. The paper \cite{Kim2026} also considers the out-of-band energy constraint, a constraint that is especially useful when designing a pulse shape for multi-user scenarios. 

In all these works, either an additive white Gaussian noise (AWGN) channel is assumed or the channel is perfectly known at the receiver. This is, however, a strong assumption for an HRLLC scenario. Assuming channel estimation is performed using pilot symbols, as is common in practical systems, a fraction of the already limited channel uses in the short-blocklength regime must be allocated to pilot transmission, thereby severely reducing the data rate. Therefore, the optimal ratio of pilot symbols to data symbols, reflecting the tradeoff between channel estimation accuracy and data rate, must be characterized to obtain a realistic benchmark for HRLLC systems.  

\paragraph*{Contributions}
In this paper, we consider the transmission of short packets via FTN signaling over a SISO memoryless block-fading channel. Specifically, we investigate a pilot-assisted FTN scheme, where both pilot and data symbols are transmitted using FTN signaling. We develop a channel estimation algorithm for the received fading blocks and derive the RCUs bound for this system. Furthermore, we explore the use of guard intervals (null transmissions) between pilot and data symbols to mitigate or entirely eliminate mutual interference. To provide design insights for HRLLC-FTN systems, we report numerical results demonstrating the utility of this framework. Our results show that, compared to Nyquist signaling, FTN requires up to $2 \dB$ less SNR to achieve a packet error probability of $10^{-5}$ in the short-blocklength regime when imperfect, pilot-assisted channel estimation is used.
\paragraph*{Notation} 
We denote random vectors and random scalars by upper-case boldface letters such as
$\rvecx$ and upper-case standard letters,
such as $\rndx$, respectively. 
Their realizations are indicated by lower-case letters of the same font. 
We use upper-case sans-serif letters $\matX$ to denote deterministic matrices. To avoid ambiguities, we use roman letters, such as $\rate$ for rate, to denote constants that are typically capitalized in the literature. The circularly-symmetric Gaussian distribution is denoted by $\jpg(0, \sigma^2)$, where $\sigma^2$ denotes the variance. The superscripts $(\cdot)^T$, $(\cdot)^H$, and $(\cdot)^*$ denotes transposition, Hermitian transposition, and complex conjugation, respectively. We finally write $\log(\cdot)$ 
to denote the natural logarithm, $\matI$ for the identity matrix, $\Prob[\cdot]$ for the probability of an event and $\Exop[\cdot]$ for the expectation operator.

\section{System Model}

\subsection{System Model}
We consider a memoryless block-fading channel, where the channel is assumed to remain constant over a block of $\nc$ channel uses and changes independently across blocks. Each transmitted packet spans $\nb$ such fading blocks, providing $\nb$ diversity branches. Consequently, each packet consists of a total of $\nb\nc$ complex-valued symbols. We assume that for each fading block, a block-wise independent complex baseband FTN signal is transmitted, and these signals are received without experiencing interference from adjacent blocks.

Let $p(t)$ be a real-valued, root-raised cosine pulse with roll-off factor $\beta$  and signaling period $\mT$. The complex baseband FTN signal transmitted over the $\ell$th diversity branch is given by
\begin{equation}
    \rndx_{\ell}(t) = \sum_{k=0}^{\nc - 1} \rndx_{k,\ell} p(t-k\delta\mT)
\end{equation}
where $\ell \in \{1,\ldots,\nb\}$ denotes the fading block index, and $\rndx_{k,\ell}$ is the complex symbol to be transmitted in the $k$th channel use over the $l$th diversity branch. The pulse shape $p(t)$ with the acceleration factor $\delta \in (0,1]$, is projected onto a basis set  $\{\phi(t-m\delta \mT)\}_m$, $m \in \mathbb{Z}$ as 
\begin{equation}
    p(t) =  \sum_{m=-\infty}^{\infty} g_{m} \phi(t - m\delta\mT)
\end{equation}
where the coefficients $g_m$ are defined as
\begin{equation}
    g_m = \int_{-\infty}^{\infty} p(t) \phi(t - m \delta \mT) \dv t.
\end{equation}
The orthonormal base signal $\phi(t)$ needs to be constant over the bandwidth $1/(\delta \mT)$, to ensure the orthogonality of orthonormal basis signals over $\delta \mT$. Then, we can state the transmitted signal as
\begin{equation}
    \rndx_{\ell}(t) = \sum_{k=0}^{\nc - 1} \sum_{m = -\infty}^{\infty} \rndx_{k,\ell} g_m \phi(t - (m+k)\delta \mT).
\end{equation}

After the transmitted signal passes through the SISO block-fading channel, the received signal for the $\ell$th fading block is expressed as
\begin{equation}
    \rndy_{\ell}(t) = \rndh_{\ell} \rndx_{\ell}(t) + \rndw_{\ell}(t).
\end{equation}
Here $\rndw_{\ell}(t)$ is the zero-mean complex Gaussian noise. The received signal is then correlated with a set of receiver filters $\phi^*(t-j\delta\mT)$, yielding the filter outputs 
\begin{align}
    \rndy_{j,\ell} &= \int \rndy_{\ell}(t) \phi^{*}(t-j\delta\mT) \dv t \\
     &= \rndh_{\ell} \sum_{k=0}^{\nc-1} \rndx_{k,\ell} g_{j-k} + \rndw_{j,\ell},
\end{align}
where the noise covariance after the correlation receiver can be found as
\begin{align}
    \nonumber
    &\Ex{}{\rndw_{j,\ell} \ltrp{\rndw_{i,\ell}}^*} \\
    &= \Ex{}{\int \rndw_{\ell}(t) \phi^*(t-j\delta\mT) \ltrp{\rndw_{\ell}}^*(u) \phi(u - i\delta\mT) \dv t \dv u} \\
    &= \mN_0 \int{\phi^*(t-j\delta\mT) \phi(t-i\delta\mT) \dv t} \\ 
    &= \begin{cases} 
        N_0, & j = i \\
        0, & j \neq i 
        \end{cases}
\end{align}
This result indicates that the noise at the receiver remains white regardless of the acceleration factor, which is expected since the received signal is projected onto an orthonormal basis.




\subsection{Channel Estimation and Decoding}
For decoding, although all samples of $\rndy_{j,\ell}$ for $j \in \mathbb{Z}$ are theoretically required, such an approach leads to an intractable model. Therefore, we assume that $g_m = 0$ for $\abs{m} > \mM$. Consequently, only $\nc + 2\mM$ samples are required for decoding. The received signal vector for the $\ell$th fading block is then given by
\begin{equation}
  \rvecy_{\ell} = \rndh_{\ell} \matG \rvecx_{\ell} + \rvecw_{\ell}  
\end{equation}
where $\rvecx_{\ell} \in \mathbb{C}^{\nc}$ is the transmitted complex symbols, $\rvecy_{\ell} \in \mathbb{C}^{\nc+2\mM}$ is the received signal, $\rndh_{\ell}$ is the fading gain and assumed to be independent and identically distributed (\iid) $\jpg(0,1)$, $\rvecw_{\ell} \in \mathbb{C}^{\nc +2\mM}$ is the noise with distribution $\jpg(0,N_0 \matI)$, $\matG \in \mathbb{C}^{(\nc + 2\mM) \times \nc}$ is a symmetric Toeplitz matrix defined as
\begin{equation}
    \matG = 
    \begin{bmatrix}
     g_{-\mM} & g_{-\mM-1} & \ldots & g_{-\nc-\mM+1}    \\ 
     g_{-\mM+1} & g_{-\mM} & \ldots & g_{-\nc-\mM+2} \\
     \vdots & \vdots & \vdots & \vdots \\
    g_{\nc+\mM-1} & g_{\nc+\mM} & \ldots & g_{\mM}
    \end{bmatrix}.
\end{equation}

For each fading block, we transmit $\np$ pilot symbols at the beginning of the signal, followed by $\ng$ guard symbols, where $0 \leq \ng \leq 2\mM$. In our setup, a guard symbol interval corresponds to a null transmission, i.e., $\vecx^{(g)} = \mathbf{0}$. Furthermore, the cases $\ng = 0$ and $\ng = 2\mM$ correspond to scenarios where no guard interval is introduced between pilot and data symbols, and where the pilot and data symbols do not interfere with each other, respectively.

To model the channel estimation, we let $\rvecx_\ell~=~[\vecx_{\ell}^{(p)}, \vecx_{\ell}^{(g)}, \rvecx_{\ell}^{(d)}]$, where $\vecx_{\ell}^{(p)} \in \mathbb{C}^{\np}$, $\vecx_{\ell}^{(g)} \in \mathbb{C}^{\ng}$, and $\rvecx_{\ell} \in \mathbb{C}^{\nd}$ denote the transmitted pilot, guard, and data symbols\footnote{The data symbols are represented with capital letters as they are stochastic from the receivers perspective, while pilot and guard symbols are not capitalized as they are always deterministic.}, respectively.  Accordingly, the pilot transmission is denoted as
\begin{align}
    \rvecy^{(p)}_{\ell} &= \rndh_\ell \matG_{p}  \vecx_{\ell}^{(p)} + \rndh_{\ell} \matG_{pg} \vecx^{(g)}_{\ell} + \rndh_\ell \matG_{pd} \rvecx^{(d)}_{\ell}  + \rvecw_{\ell}^{(p)} \\
    &= \rndh_{\ell} \matG_{p} \vecx^{(p)}_{\ell} + \rndh_{\ell} \matG_{pd} \rvecx_{\ell}^{(d)}  + \rvecw_{\ell}^{(p)}
    \label{eq:chEst_eq2}
\end{align}
where $\rvecy^{(p)}_{\ell} \in \mathbb{C}^{\np + 2\mM}$ denotes the received vector that contains the transmitted pilot symbols, $\rvecw_{\ell}^{(p)} \in \mathbb{C}^{\np}$ is the vector containing noise samples, $\matG_p  \in \mathbb{C}^{(\np + 2\mM) \times \np} $, $\matG_{pg} \in \mathbb{C}^{(\np + 2\mM) \times \ng}$, $\matG_{pd}~ \in~\mathbb{C}^{(\np + 2\mM) \times \nd}$ are sub-matrices of $\matG$ as 
\begin{equation}
    \label{eq:Gmat_pilot}
    \matG = \begin{bmatrix}
        \matG_{p} & \matG_{pg} & \matG_{pd} \\ 
        \tilde{\matG}_{pd} & \tilde{\matG}_{pg} & \tilde{\matG}_{p}
    \end{bmatrix}
\end{equation}
$\nd = \nc - (\np + \ng)$ is the number of data symbols transmitted over a fading block. Since the data symbols are unknown at the receiver, the ISI term $\rndh \matG_{pd}\rvecx_{\ell}^{(d)}$ in \eqref{eq:chEst_eq2} cannot be removed. Under this condition, the best linear unbiased estimator (BLUE) for the channel gain $\rndh_{\ell}$ is given by
\begin{equation}
    \hat{\rndh}_{\ell} = \ltrp{\ltrp{\matG_{p}\vecx_{\ell}^{(p)}}^{H} \rvecy_{\ell}^{(p)}}  \ltrp{\ltrp{\matG_{p} \vecx_{\ell}^{(p)}}^{H}  \ltrp{\matG_{p} \vecx_{\ell}^{(p)}}}^{-1} .
\end{equation}

Within each block, the pilot-transmission phase is followed by a guard interval and a data-transmission phase consisting of $\nd$ complex symbols per block. We assume that the $\nb\nd$ symbols are selected from a codebook $\setC$ of size $\tilde{\mM}~=~\ceil{\exp(\nb\nc\rate)}$, where $\rate$ denotes the transmission rate in nats per channel use.

To state the input-output relation of the data phase, similar to \eqref{eq:Gmat_pilot}, we first divide $\matG$ to 6 sub-matrices as 
\begin{equation}
    \matG = \begin{bmatrix}
        \tilde{\matG}_d & \tilde{\matG}_{dg} & \tilde{\matG}_{dp} \\
        \matG_{dp} & \matG_{dg} & \matG_d
    \end{bmatrix}
\end{equation}
where $\matG_d \in \mathbb{C}^{(\nd+2\mM) \times \nd}$, 
$\matG_{dg} \in \mathbb{C}^{(\nd + 2\mM) \times \ng}$
$\matG_{dp} \in \mathbb{C}^{(\nd + 2\mM) \times \np}$.
Using these sub-matrices, the received vector can be stated as
\begin{align}
    \rvecy_{\ell}^{(d)} &= \rndh_{\ell} \matG_{d}  \rvecx_{\ell}^{(d)} + \rndh_{\ell} \matG_{dg} \vecx_{\ell}^{(g)} + \rndh_{\ell} \matG_{dp}  \vecx_{\ell}^{(p)} + \rvecw_{\ell}^{(d)} \\
    &= \rndh_{\ell} \matG_{d} \rvecx_{\ell}^{(d)} + \rndh_{\ell}\vecz_{\ell}^{(d)} + \rvecw_{\ell}^{(d)}
    \label{eq:dataSignal_form1}
\end{align}
where
\begin{equation}
    \vecz_{\ell}^{(d)} = \matG_{dp} \vecx_{\ell}^{(p)}.
\end{equation}
The relationship in \eqref{eq:dataSignal_form1} can be simplified by utilizing the singular value decomposition (SVD) of $\matG_d = \matU \Lambda \matV^{H}$, where $\matU \in \mathbb{C}^{(\nd+2\mM) \times (\nd + 2\mM)}$ and $\matV \in \mathbb{C}^{\nd \times \nd}$ are unitary matrices, and $\Lambda \in \mathbb{R}^{(\nd + 2\mM) \times \nd}$ is a diagonal matrix with the singular values $\{\lambda_0, \ldots, \lambda_{\nd-1}\}$ of $\matG_d$ on its main diagonal. Multiplying both sides of \eqref{eq:dataSignal_form1} by $\matU^H$, the system can be equivalently expressed as\footnote{Indeed, multiplying a random vector by a unitary matrix preserves its average power.}
\begin{equation}
\label{eq:dataSignal_form2}
    \tilde{\rndy}_{k,\ell}^{(d)} = \rndh_{\ell} \lambda_k \tilde{\rndx}_{k,\ell}^{(d)} + \rndh_{\ell} \tilde{z}_{k,\ell}^{(d)}  + \tilde{\rndw}_{k,\ell}^{(d)}
\end{equation}
where $\tilde{\rvecx}_{\ell}^{(d)} = \matV^H \rvecx_{\ell}^{(d)}$, $\tilde{\rvecw}_{\ell}^{(d)} = \matU^{H} \rvecw_{\ell}^{(d)}$,
\begin{equation}
    \tilde{\vecz}_{\ell}^{(d)} = \matU^H \matG_{dp} \vecx^{(p)}_{\ell} 
\end{equation}
and $\tilde{\rndx}^{(d)}_{k,\ell}$, $\tilde{\rndy}^{(d)}_{k,\ell}$, $\tilde{z}^{(d)}_{k,\ell}$, $\tilde{\rndw}^{(d)}_{k,\ell}$ are the $k$th element of $\tilde{\rvecx}^{(d)}_{\ell}$, $\tilde{\rvecy}^{(d)}_{\ell}$, $\tilde{\vecz}^{(d)}_{\ell}$ and $\tilde{\rvecw}^{(d)}_{\ell}$, respectively. The expression in \eqref{eq:dataSignal_form1} can be interpreted as $\nd$ parallel AWGN channels, each characterized by a channel gain $\rndh_{\ell}$, a distortion term $\tilde{z}_{k,\ell}^{(d)}$, and an SNR that scales with $\lambda_k^2$. Note that if the receiver has access to perfect channel state information, the distortion term can be completely removed.


To perform decoding, the receiver seeks the codeword in the codebook that is closest to the received
signal, once each part of the codeword corresponding to a different fading block is scaled by the available channel estimate.
Mathematically, given the received vector and the channel estimates the decoded codeword is determined as
\begin{equation}
    \tilde{\hat{\vecx}}^{(d)} = \argmin_{ \bar{\vecx}_{\ell}^{(d)}  \in f(\setC) \forall \ell }  \sum_{\ell=1}^{\nb} \vecnorm{\tilde{\vecy}_{\ell}^{(d)}
  -(\hat{h}_{\ell}\Lambda\bar{\vecx}_{\ell}^{(d)} + \hat{h}_{\ell}\tilde{\vecz}^{(d)}_{\ell})}^{2} \label{eq:snn_dec}
\end{equation}
where $f(\vecx^{(d)}) = [(\matV^H \vecx_{1}^{(d)})^{T}, \ldots, (\matV^H \vecx_{\nb}^{(d)})^{T} ]^{T} $ and $f(\setC)$ is the image of $\setC$ according to the definition of the function.

This decoder, known as the mismatched scaled-nearest neighbor (SNN) decoder \cite{Kislal2024}, coincides with the maximum likelihood (ML) decoder only in the presence of perfect channel-state information, i.e., $\hat{h}_\ell = h_\ell$ for $\ell = 1, \dots, \nb$.

\subsection{Power Budget and Allocation}
The transmit power of an FTN signal for each fading block can be found as \cite[Sec. III]{Zhang2025Mar}
\begin{align}
    \mP_{\text{Tx},\ell} &= \Ex{}{\frac{1}{\nc \delta \mT} \int_{-\infty}^{\infty} \abs{\rndx_{\ell}(t)}^2 \dv t } \\ 
    &= \frac{1}{\nc \delta \mT} \text{tr}\ltrp{\matG^{H} \matG \Ex{}{\rvecx_{\ell} \rvecx_{\ell}^H}}.
    \end{align}
Here, we observe that the acceleration factor and the ISI matrix $\matG$ directly affect the transmit power. Consequently, for a fair comparison with Nyquist signaling (i.e., $\delta = 1$), this impact must be accounted for. To this end, we constrain the power of the transmitted symbols\footnote{As we shall resort to random coding in the next section for the finite-blocklength achievability bounds, the instantaneous transmit power of a symbol or the power of each specific codeword cannot be strictly bounded. Instead, we bound the average transmit power, averaged over the randomly constructed codebook ensemble.} such that $\mP_{\text{Tx},\ell} \leq \mP$.

Next, we discuss the power allocation for the transmitted symbols. Let $\vecp^{(p)} = [\rho_1^{(p)}, \ldots, \rho_{\np}^{(p)}]^T$ denote the $\np$-dimensional pilot power vector, where $\rho_i^{(p)}$ is the transmit power of the $i$th pilot symbol for $i \in \{1, \ldots, \np\}$. Similarly, let $\vecp^{(d)} = [\rho_1^{(d)}, \ldots, \rho_{\nd}^{(d)}]^T$ be the $\nd$-dimensional vector where $\tilde{\rho}_k$ represents the transmit power of the $k$th transformed data symbol $\tilde{\rndx}_{k,\ell}$  for $j \in \{1, \ldots, \nd\}$. In this study, we focus on the power allocation of the data symbols and assume a uniform power allocation for the pilot symbols, given by $\rho_i^{(p)} = \mP$ for all $i \in \{1, \ldots, \np\}$, i.e., the pilot symbols carry a total energy of $\mathrm{E}_{\np} = \mP \delta \mT \np $, while guard symbols carry no energy.

For the data symbols, we may allocate power based on the gains $\lambda_k$ of the parallel channels in \eqref{eq:dataSignal_form2} using the classic water-filling solution as\footnote{Note that while water-filling maximizes capacity, it does not necessarily maximize the rate for a given packet error probability $\epsilon$ in the short-blocklength regime.}
\begin{equation}
    \rho_k^{(d)} = \left( \mu - \frac{1}{\lambda_k^{2}} \right)^{+}
\end{equation}
where $\mu$ is the water-level (cut-off threshold). The value of $\mu$ can be determined iteratively by incrementing it in small steps until the power constraint $\mP_{\text{Tx},\ell} \leq \mP$ is satisfied.

\section{A Non-Asymptotic Achievability Bound}
Similar to most achievability results in information theory, the achievability bounds in this paper are obtained through a random-coding argument. Specifically, we evaluate the average error probability averaged over a randomly constructed ensemble of codebooks. We consider the \iid Gaussian ensemble, in which each symbol of each codeword is drawn independently from $\jpg(0,1)$ and then scaled by $\sqrt{\rho_{k}^{(d)}}$ for $k \in \{1,\ldots,\nd\}$ prior to transmission. Although the Gaussian ensemble is not strictly optimal in the finite-blocklength regime, it is utilized extensively in the literature as it leads to tractable expressions when applied to mismatched SNN decoding. 
\subsection{RCU \& RCUs Bounds}
The RCU and RCUs bounds were introduced in \cite{Scarlett2014} for mismatched decoding. For completeness, we present both bounds along with their respective derivations. In our derivations, we omit the superscript $(d)$ for brevity.

Let $i$th codeword in the codebook is  $\rvecx^{(i)}$. 
Mathematically, the average error probability can be upper bounded as
\begin{align}
    \nonumber
    &\epsilon(\rvecx^{(1)},\ldots,\rvecx^{(\tilde{\mM})}) \\
    & \quad  \leq  \frac{1}{\tilde{\mM}} \sum_{i=1}^{\tilde{\mM}} \Prob\ltrsqr{\Union_{j=1,j\neq i}^{\tilde{\mM}} P_{\rvecy|\rvecx}(\rvecy|\rvecx^{(j)}) \geq P_{\rvecy|\rvecx}(\rvecy|\rvecx^{(i)}) }  
    \label{eq:RCU_1}
\end{align}
where we treat ties between $P_{\rvecy| \rvecx}(\rvecy|\rvecx^{(j)})$ and $P_{\rvecy|\rvecx}(\rvecy|\rvecx^{(i)})$ as errors. These ties occur very rarely, thus the bound in \eqref{eq:RCU_1} is tight. We next consider the error probability averaged over the entire codebook ensemble. 
Since all codewords are \iid, we may assume that $\rvecx^{(1)}$ is transmitted without loss of generality. Consequently, the average error probability $\epsilon = \Ex{}{\epsilon(\rvecx^{(1)}, \dots, \rvecx^{(\tilde{\mM})})}$ is computed as
\begin{align}
    \epsilon &\leq \Prob\ltrsqr{\Union_{j=2}^{\tilde{\mM}} P_{\rvecy|\rvecx}(\rvecy|\rvecx^{(j)}) \geq P_{\rvecy|\rvecx}(\rvecy|\rvecx^{(1)})} \\ 
             &\leq \Exop\Biggl[\min\Biggl\{1, \sum_{j=2}^{\tilde{\mM}} \Prob\bigl[P_{\rvecy|\rvecx}(\rvecy|\rvecx^{(j)}) \nonumber\\ 
             & \qquad \qquad \qquad \qquad \geq P_{\rvecy|\rvecx}(\rvecy|\rvecx^{(1)}) 
             \given \rvecx^{(1)}, \rvecy \bigr]  \Biggr\} \Biggr]\label{ch2_eq:RCU_step3} \\ 
             &= \Exop\bigl[\min\bigl\{1, (\tilde{\mM}-1) \Prob\bigl[P_{\rvecy|\rvecx}(\rvecy|\bar{\rvecx}) \nonumber \\
             & \qquad \qquad \qquad \qquad \geq P_{\rvecy|\rvecx}(\rvecy|\rvecx) \given \rvecx, \rvecy \bigr] \bigr\}\big]
             \label{ch2_eq:RCU_step4}
\end{align}
where for notational brevity, in \eqref{ch2_eq:RCU_step4} we do not explicitly denote the dependence of $\rvecx$ on $i$ due to the random coding construction, each codeword follows the same distribution as $\rvecx$. 
\begin{equation}
    P_{\rvecx,\rvecy,\bar{\rvecx}}(\rvecx,\rvecy,\bar{\rvecx}) = P_{\rvecx}(\rvecx) P_{\rvecy|\rvecx}(\rvecy|\rvecx) P_{\rvecx}(\bar{\rvecx})
\end{equation}
which indicates that $\bar{\rvecx}$ has the same distribution as $\rvecx$, and it is independent of $\rvecx$ and $\rvecy$. 
In \eqref{ch2_eq:RCU_step3} we condition on $\rvecx^{(1)}$ and $\rvecy$ yields a tighter bound prior to applying the union bound. Furthermore, \eqref{ch2_eq:RCU_step4} results from the \iid nature of the codebook, where the symmetry of the construction ensures that the $(\tilde{\mM}-1)$ probability terms are identical. RCU bound, although tightest achievability bound available in the literature, it is also very hard to evaluate numerically. The probability inside the expectation in \eqref{ch2_eq:RCU_step4} needs to be evaluated with a very high resolution as it scales with $\tilde{\mM}-1$. For instance, even when the blocklength is as small as $n=150$ and $\rate = 2/3$ nats per channel use, the codebook size is $\tilde{\mM} \geq 10^{15}$. 

We may apply the Markov inequality to relax the RCU bound and obtain the RCUs bound as
\begin{align}
    \epsilon &\leq \Ex{}{\min\ltrcurley{1, (\tilde{\mM}-1) \frac{ \Ex{\bar{\rvecx}}{P_{\rvecy|\rvecx}(\rvecy|\bar{\rvecx})^s} } {P_{\rvecy|\rvecx}(\rvecy|\rvecx)^{s}} } } \label{ch2_eq:RCUs_part1} \\ 
    &= \Ex{}{e^{-\ltrp{\infden(\rvecx;\rvecy) - \log(\tilde{\mM}-1)}^{+}}} \label{ch2_eq:RCUs_Expect} \\
    &= \Ex{}{\Ex{}{e^{-\ltrp{\sum_{\ell=1}^{\nb}\sum_{k=1}^{\nd}\infden(\rndx_{k,\ell};\rndy_{k,\ell}) - \log(\tilde{\mM}-1)}^{+}} \ggiven \rvech, \hat{\rvech}}  } \label{eq:RCUs_hExpect}
\end{align}
where 
\begin{equation}
    \label{ch2_eq:infoDens}
    \infden(\vecx;\vecy) = \log \frac{P_{\rvecy|\rvecx}(\vecy|\vecx)^{s}}{\Ex{\bar{\rvecx}}{P_{\rvecy|\rvecx}(\vecy|\bar{\rvecx})^s} }.
\end{equation}
is the generalized information density and $s > 0$ is an optimization parameter for which the bound holds. Equation \eqref{eq:RCUs_hExpect} follows from the fact that, under the assumed random coding, every transmitted symbol $x_{k,\ell}$ is independent; consequently, each received symbol of a codeword is conditionally independent given the channel gain and its estimate. This concludes our derivation of RCUs bounds. 

To evaluate the RCUs bound for the mismatched SNN decoder \eqref{eq:snn_dec}, one needs to evaluate $\infden(\tilde{x}_{k,\ell}^{(d)}, \tilde{y}^{(d)}_{k,\ell})$ which can be found in closed-form for this system model as
\begin{align}
    &\infden(\tilde{x}^{(d)}_{k,l};\tilde{y}^{(d)}_{k,l}) = \log \frac{e^{-s \abs{\tilde{y}^{(d)}_{k,l} - \hat{h}_{\ell}(\lambda_k\tilde{x}^{(d)}_{k,\ell}+\tilde{z}^{(d)}_{k,\ell})}^2}}{\Ex{\bar{\rndx}}{e^{-s \abs{\tilde{y}^{(d)}_{k,\ell} - \hat{h}_\ell(\lambda_k\bar{\rndx} + \tilde{z}^{(d)}_{k,\ell} )}^2}}}  \\
    &= \log\ltrp{1+s \rho_{k}^{(d)} \abs{\hat{h}_{\ell}}^2} -s \abs{\tilde{c}^{(d)}_{k,\ell}}^2  + \frac{s\abs{\tilde{y}^{(d)}_{k,\ell} - \hat{h}_{\ell}\tilde{z}^{(d)}_{k,\ell}}^2}{1+s\abs{\hat{h}_{\ell}}^2\rho_k^{(d)}}
\end{align}
where 
\begin{equation}
    \tilde{c}^{(d)}_{k,\ell} = \tilde{y}^{(d)}_{k,\ell} - \hat{h}_{\ell}\ltrp{\lambda_k\tilde{x}^{(d)}_{k,\ell} + \tilde{z}^{(d)}_{k,\ell}}.
\end{equation}

\section{Numerical Results and Discussion}
In this section, a set of simulation results are presented to assess the impact of FTN signaling in the short-blocklength regime. In particular, we aim to answer the following questions:
\begin{itemize}
\item How significantly does power optimization affect FTN performance in the short-blocklength regime? Is it essential, or can it be neglected?
\item How does the number of fading blocks affect FTN signaling performance compared to Nyquist signaling?
\item How does the number of pilot and guard symbols affect performance?
\end{itemize}
\begin{figure}[t]
    \centering
    \includegraphics[width=0.88\columnwidth]{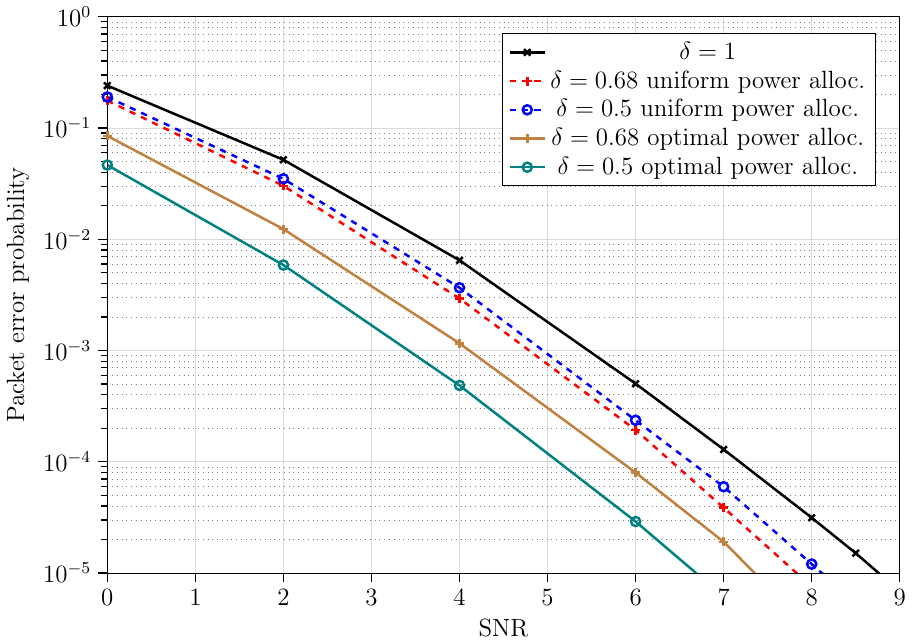}
    \caption{Packet error probability as a function of SNR. Here, $\nb =8$, $\rate_{b/s/h} = 0.233$ bit per second per Hertz, $\nb\nc=288$; $\np$, $\ng$ and $s$ are optimized.} 
    \label{fig:results_set1_bonus}
\end{figure}
We consider a Rayleigh-fading scenario where $\rndh_{\ell}$ 
are i.i.d. with $\jpg(0,1)$ for all $\ell \in \{1,\ldots, \nb\}$. We set $\beta = 0.5$, $\mM = 10$, the system SNR as 
$\mP/\mN_0$, and the acceleration factor as $\delta \in \{0.5, 0.68, 1\}$.\footnote{The choice of $\delta = 0.68$ comes from the theoretical limit  $1/(1+\beta)$ which is the acceleration factor before exceeding the bandwitdh capability of the pulse.} We also set a blocklength of $\nb\nc = 288$ and a spectral efficiency $\rate_{b/s/h} = 0.233$ bit per second per Hertz. The rate in nats per channel use can be found by $\rate = \rate_{b/s/h} (1+\beta) \delta \log(2)$. All reported error probabilities are obtained via the RCUs bound in \eqref{eq:RCUs_hExpect}, where the expectations are evaluated using Monte Carlo simulations and the parameter $s$ is optimized numerically.

In Fig. \ref{fig:results_set1_bonus}, we report the packet error probability as a function of SNR for $\nb=8$ to examine how FTN and Nyquist signaling behave across different error regimes. We observe that FTN signaling consistently requires approximately $2 \dB$ less SNR to achieve the same error probability as Nyquist signaling ($\delta = 1$) for $10^{-1} \leq \epsilon \leq 10^{-5}$. We also observe that uniform power allocation reduces this gain about $1$ dB.

\begin{figure}[t]
    \centering
    \includegraphics[width=0.88\columnwidth]{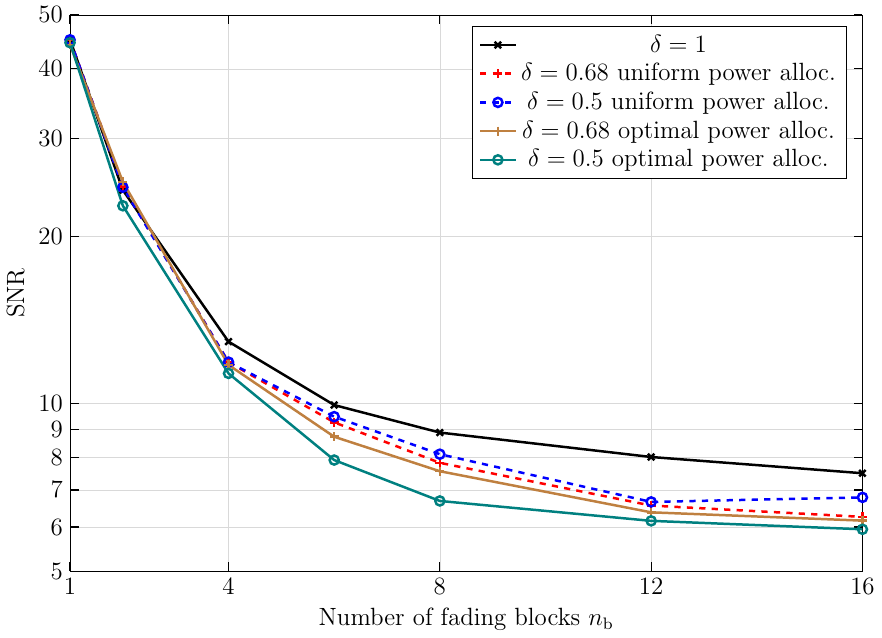}
    \caption{Required minimum SNR to achieve packet error probability $\epsilon \leq 10^{-5}$ as a function of $\nb$. Here, $\nb\nc=288$, $\rate_{b/s/h} = 0.233$ bit per second per Hertz ; $\np$, $\ng$ and $s$ are optimized.} 
    \label{fig:results_set1}
\end{figure}

In Fig. \ref{fig:results_set1}, we illustrate the minimum $\mP$ required to achieve a packet error probability of $\epsilon = 10^{-5}$ as a function of the number of fading blocks $n_b$. For each data point, the parameters $n_p$ and $n_g$ are optimized. We observe that as $n_b$ increases, the required SNR decreases across all settings, and FTN signaling, particularly when optimal power allocation is used, provides up to a $2 \dB$ gain over Nyquist signaling for $n_b > 1$. In the absence of diversity, i.e. $\nb = 1$, the packet error probability is dominated by deep-fading events, and FTN and Nyquist signalling have the same performance.

\begin{figure}[t]
    \centering
    \includegraphics[width=0.88\columnwidth]{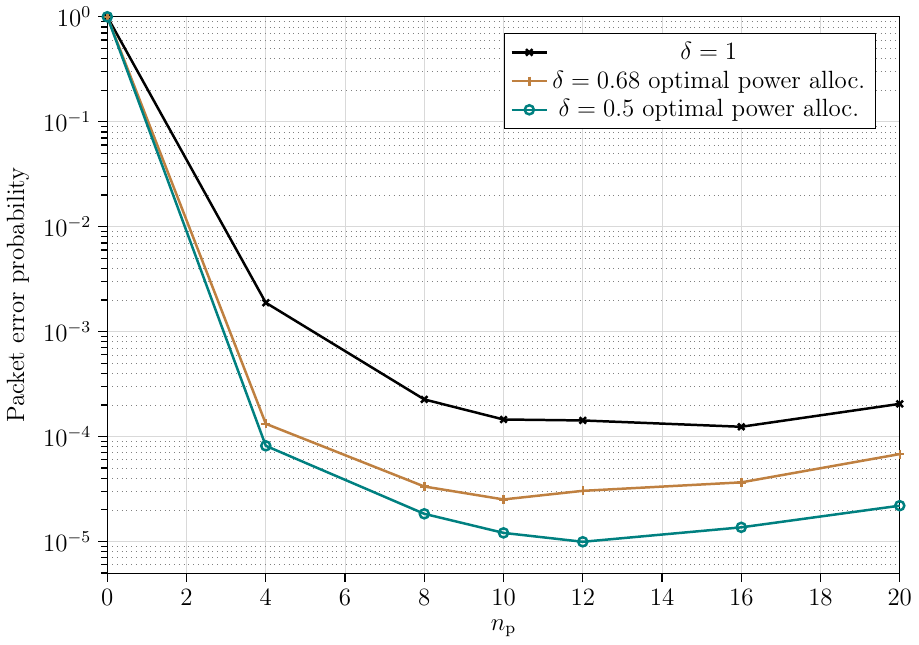}
    \caption{Packet error probability as a function of $\np$. Here, SNR is $8 \dB$, $\nb\nc=288$, $\rate = 0.233$ bit per Hertz per second; $\ng$ and $s$ are optimized. } 
    \label{fig:results_set2}
\end{figure}

In Fig. \ref{fig:results_set2}, we report the packet error probability as a function of the number of pilot symbols $\np$ for $\nb = 6$ and optimal power allocation, where $\ng$ is optimized for each $\np$. We observe that optimizing $\np$ during system design is crucial for both FTN and Nyquist signaling to achieve peak performance. Indeed, when $\np$ is too large, the remaining data symbols $\nd$ are reduced; to maintain the same $\rate$, a weaker channel code must be employed. Conversely, when $\np$ is too small, the channel estimation quality suffers, which severely degrades the performance.

\begin{figure}[t]
    \centering
    \includegraphics[width=0.88\columnwidth]{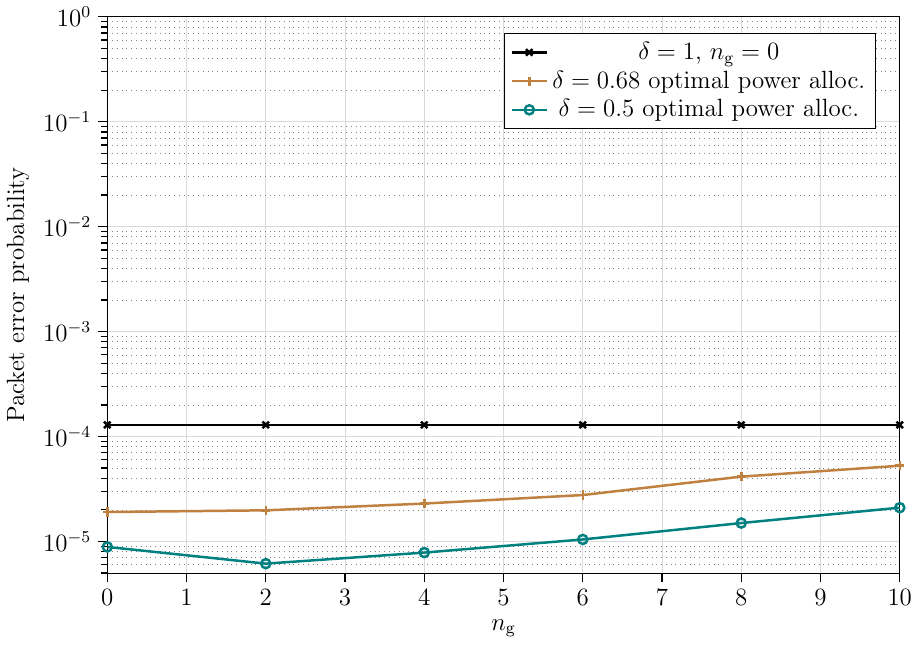}
    \caption{Packet error probability as a function of $\np$. Here, SNR is $7 \dB$, $\nb\nc=288$, $\rate = 0.233$ bit per second per Hertz; $\ng$ and $s$ are optimized. } 
    \label{fig:results_set3}
\end{figure}

Finally, Fig. \ref{fig:results_set3} shows the packet error probability as a function of the number of guard symbols $\ng$ for $\nb = 8$ and optimal power allocation, where $\np$ is optimized for each $\ng$. Interestingly, we observe that the inclusion of guard symbols improves performance only slightly for $\ng = 2$ when optimal transmit power allocation is employed; beyond this specific case, increasing $\ng$ decreases performance across all other configurations\footnote{Although not reported in this paper, we have not observed any gain for $\ng > 2$ at other values of $\nb$ for $\epsilon \geq 10^{-5}$.}. In other words, unlike the optimization of pilot symbols $\np$, when designing a communication system with FTN signaling, one may either completely omit guard symbols for simplicity or set $\ng = 2$ to achieve a marginal gain.

\section{Conclusion \& Future Work}
In this paper, we investigated faster-than-Nyquist (FTN) signaling for HRLLC by deriving a non-asymptotic achievability bound that accounts for pilot-assisted channel estimation overhead. Our numerical results show that FTN provides up to a $2 \dB$ SNR gain over Nyquist systems, provided that power allocation and pilot overhead are optimized. While guard symbols offer only marginal benefits, precise tuning of the pilot quantity is essential to balance estimation accuracy and data rate. In our future work, we will expand the finite-blocklength analysis from single-input single-output to multiple-input multiple-output systems to further exploit spatial diversity and multiplexing gains in HRLLC scenarios. We also plan to investigate the impact of having different acceleration factors for pilot and data transmission.  
\bibliographystyle{IEEEtran}
\bibliography{FTN}

@article{durisi16-09a,
	author = {Giuseppe Durisi and Tobias Koch and Petar Popovski},
	journal = {Proc. {IEEE}},
	month = sep,
	number = {9},
	pages = {1711--1726},
	title = {Towards Massive, Ultra-Reliable, and Low-Latency Wireless Communication with Short Packets},
	volume = {104},
	year = {2016}}

@ARTICLE{Mazo1975,
  author={Mazo, J. E.},
  journal={The Bell System Technical Journal}, 
  title={Faster-than-{N}yquist signaling}, 
  year={1975},
  volume={54},
  number={8},
  pages={1451-1462},
  doi={10.1002/j.1538-7305.1975.tb02043.x}}

@ARTICLE{Ganji2020,
  author={Ganji, Mehdi and Zou, Xun and Jafarkhani, Hamid},
  journal={IEEE Commun. Lett.}, 
  title={On the Capacity of Faster Than {N}yquist Signaling}, 
  year={2020},
  volume={24},
  number={6},
  pages={1197-1201},
  doi={10.1109/LCOMM.2020.2980263}}

@ARTICLE{Zhang2023,
  author={Zhang, Zichao and Yuksel, Melda and Guvensen, Gokhan M. and Yanikomeroglu, Halim},
  journal={IEEE Commun. Lett.}, 
  title={Capacity Region of Asynchronous Multiple Access Channels With {FTN}}, 
  year={2023},
  volume={27},
  number={7},
  pages={1719-1723},
  doi={10.1109/LCOMM.2023.3269739}}

@ARTICLE{Zichao2023,
  author={Zhang, Zichao and Yuksel, Melda and Yanikomeroglu, Halim},
  journal={IEEE Trans. Wireless Commun.}, 
  title={Faster-Than-{N}yquist Signaling for {MIMO} Communications}, 
  year={2023},
  volume={22},
  number={4},
  pages={2379-2392},
  doi={10.1109/TWC.2022.3211327}}

@ARTICLE{Johan2019,
  author={Östman, Johan and Durisi, Giuseppe and Ström, Erik G. and Coşkun, Mustafa C. and Liva, Gianluigi},
  journal={IEEE Trans. Commun.}, 
  title={Short Packets Over Block-Memoryless Fading Channels: Pilot-Assisted or Noncoherent Transmission?}, 
  year={2019},
  volume={67},
  number={2},
  pages={1521-1536},
  doi={10.1109/TCOMM.2018.2874993}}

@InProceedings{Lin2026,
author="Lin, Pingping
and Zhang, Zhirong",
title="{URLLC} in {5G-A}: Enabling Critical Applications and Real-Time Services",
booktitle="Proc. of the International Symposium on Intelligent Computing and Networking",
year="2026",
pages="551--562"
}

@ARTICLE{Mohammadkarimi2021,
  author={Mohammadkarimi, Mostafa and Schober, Robert and Wong, Vincent W. S.},
  journal={IEEE Commun. Lett.}, 
  title={Channel Coding Rate for Finite Blocklength Faster-Than-{N}yquist Signaling}, 
  year={2021},
  volume={25},
  number={1},
  pages={64-68},
  doi={10.1109/LCOMM.2020.3021976}}

@ARTICLE{Polyanskiy2010,
  author={Polyanskiy, Yury and Poor, H. Vincent and Verdu, Sergio},
  journal={IEEE Trans. Info. Theory}, 
  title={Channel Coding Rate in the Finite Blocklength Regime}, 
  year={2010},
  volume={56},
  number={5},
  pages={2307-2359},
  doi={10.1109/TIT.2010.2043769}}

@INPROCEEDINGS{Zhang2025Mar,
  author={Zhang, Zichao and Yuksel, Melda and Yanikomeroglu, Halim and Ng, Benjamin K. and Lam, Chan–Tong},
  booktitle={Proc. Wireless Commun. Netw. Conf. (WCNC)}, 
  title={Maximum Channel Coding Rate of Finite Block Length {MIMO} Faster-Than-{Nyquist} Signaling}, 
  year={2025},
  volume={},
  number={},
  pages={01-06},
  month={Mar.}
}

@ARTICLE{Kim2026,
  author={Kim, Yong Jin Daniel},
  journal={IEEE Trans. Commun.}, 
  title={Faster-Than-{Nyquist} Signaling in the Finite Time-Bandwidth Product Regime}, 
  year={2026},
  volume={74},
  number={},
  pages={5605-5618},
  month={Feb.}
  }

@ARTICLE{Scarlett2014,
  author={Scarlett, Jonathan and Martinez, Alfonso and Fabregas, Albert Guillen i},
  journal={IEEE Trans. Inf. Theory}, 
  title={Mismatched Decoding: Error Exponents, Second-Order Rates and Saddlepoint Approximations}, 
  year={2014},
  volume={60},
  number={5},
  pages={2647-2666}}

@INPROCEEDINGS{Ince2024,
  author={Ince, Ahmet Melih and Canbilen, Ayse Elif and Yanikomeroglu, Halim},
  booktitle={Proc. International Conf. Commun., Signal Process., and Appl. (ICCSPA)}, 
  title={{HAPS}-Enabled {V2X} Architecture for Hyper Reliable and Low-Latency Communication {(HRLLC)} in {6G} Networks}, 
  year={2024},
  volume={},
  number={},
  pages={1-6}}

@ARTICLE{Liu2025,
  author={Liu, Zhaoqing and Li, Kang and Wang, Yan and Zhu, Pengcheng},
  journal={IEEE Commun. Lett.}, 
  title={Martingale-Based Scheduling of Deterministic Delay Traffic for {HRLLC} Industrial Automation}, 
  year={2025},
  volume={29},
  number={12},
  pages={2880-2884}}

@INPROCEEDINGS{Tao2023,
  author={Tao, Tao and Wang, Yang and Li, Dong and Wan, Yan and Baracca, Paolo and Wang, Ailing},
  booktitle={Proc. Veh. Technol. Conf. (VTC2023-Fall)}, 
  title={{6G} Hyper Reliable and Low-latency Communication – Requirement Analysis and Proof of Concept}, 
  year={2023},
  volume={},
  number={},
  pages={1-5}}

@ARTICLE{Yiyang2026,
  author={Li, Yiyang and Song, Xianxin and Wei, Zhiqing and Feng, Zhiyong and Xu, Jie},
  journal={IEEE Trans. Wireless Commun.}, 
  title={Joint Task Scheduling and Communication-Computation Optimization for Wireless Networked Control With {HRLLC}}, 
  year={2026},
  volume={25},
  number={},
  pages={2654-2667}}

@ARTICLE{Anderson2013,
  author={Anderson, John B. and Rusek, Fredrik and Öwall, Viktor},
  journal={Proceedings of the IEEE}, 
  title={{Faster-than-Nyquist} Signaling}, 
  year={2013},
  volume={101},
  number={8},
  pages={1817-1830}}

@ARTICLE{Kislal2024,
  author={Kislal, A. Oguz and Rajiv, Madhavi and Durisi, Giuseppe and Ström, Erik G. and Mitra, Urbashi},
  journal={IEEE Trans. Wireless Commun.}, 
  title={Is Synchronization a Bottleneck for Pilot-Assisted {URLLC} Links?}, 
  year={2024},
  volume={23},
  number={12},
  pages={17945-17958}}

\end{document}